\documentstyle[fleqn]{article}

\newlength{\dinwidth}
\newlength{\dinmargin}
\def\as{\alpha_{\mbox{\tiny S}}}
\def\alb{\bar\as}
\def\mR{\mu_{\mbox{\tiny R}}}
\def\res{\,{\mbox{\scriptsize jet}}}
\def\om{\omega}

\def\beq{\begin{equation}}
\def\beeq{\begin{eqnarray}}
\def\eeq{\end{equation}}
\def\eeeq{\end{eqnarray}}

\begin{document}

\begin{flushright}
MC-TH-98-20\\
December 1998\\
\end{flushright}
\begin{center}

{\Large \bf\boldmath Final states in small $x$ deep inelastic scattering
\footnote{Talk presented to the working groups 3 (jets and fragmentation) 
and 5 (low x and diffraction) at the third UK Phenomenology Workshop on 
HERA Physics, Durham, September 1998.}}

J.R.~Forshaw$^a$, A.~Sabio Vera$^a$ and B.R. Webber$^b$

$^a$Department of Physics and Astronomy,\\
University of Manchester,\\
Manchester, M13 9PL, England.\\
$^b$Cavendish Laboratory,\\
Madingley Road,\\
Cambridge CB3 OHE, UK.\\
\end{center}

\begin{abstract}
This talk summarises our work on the
calculation of small-$x$ jet rates within the BFKL and CCFM approaches.
The two approaches are proven to yield the same results at the leading
logarithm level to order ${\bar \alpha}_{S}^{3}$.
The proof is then extended to all orders.
\end{abstract}
{\large \bf \boldmath 1 Introduction.}

An important feature of perturbative QCD is the coherent emission of soft 
gluons in deep inelastic scattering (DIS) at small Bjorken-$x$ $~\cite{C}$. 
This is specially relevant in the calculation of exclusive quantities, e.g. 
the number of
gluons emitted in the final state. In DIS colour coherence leads to angular 
ordering with increasing opening angles towards the hard scale
(the virtuality of the photon). If $z_i$ is the fraction of energy of 
the $(i-1)th$ gluon carried off by the $i$th gluon and $q_i$ the transverse 
momentum then,
in the limit $z_{i}\ll1$, the angular ordering is implemented by the 
condition $q_{i+1}>z_{i}q_{i}$.

   When the values of  $x$ are small enough logarithms in $1/x$ need to be 
summed. This summation is performed by the 
Balitsky-Fadin-Kuraev-Lipatov (BFKL) equation which at leading order sums 
terms $\sim [\alpha_{S}\ln (1/x)]^{n}$ $ ~\cite{BFKL}$. Recently the next-
to-leading terms have
 also been computed $~\cite{NLOA,NLOB}$. 

   The ($t$=0) BFKL equation for $f_{\omega}(\mbox{\boldmath $k$})$, the 
unintegrated structure function in $\omega$-space ($\omega$ is the variable 
conjugate to $x$), expressed in a suitable form for exclusive quantities 
is~\cite{C,M}: 
$$f_{\omega}(\mbox{\boldmath $k$})=f_{\omega}^{0}(\mbox{\boldmath $k$})
+\bar\alpha_{S}\int\frac{d^{2}\mbox{\boldmath $q$}}{\pi q^{2}}\int_{0}^{1}
\frac{dz}{z}z^{\omega}\Delta_{R}(z,k)\Theta(q-\mu)f_{\omega}(\mbox{\boldmath 
$q$}+\mbox{\boldmath $k$}),$$
where $\mu$ is a collinear cutoff, \mbox{\boldmath $q$} the transverse 
momentum of the emitted gluon, and $\Delta_{R}(z_{i},k_{i})=\exp\left[-\bar
\alpha_{S}\ln\frac{1}{z_{i}}\ln
\frac{k_{i}^{2}}{\mu^2}\right],$ the gluon Regge factor which sums all the 
virtual contributions, with \(k_i\equiv|\mbox{\boldmath $k$}_{i}|\), and 
$\bar\alpha_{S}\equiv 
3\alpha_{S}/\pi$.

    This form of the BFKL equation has a kernel which, under iteration, 
generates real gluon emissions 
with all the virtual corrections summed to all orders. As such, it is 
suitable for the study of the final state. 

   Defining the structure function $F_{0\omega}(Q,\mu)$ by integrating over 
all $\mu^2\leq q_i^2 \leq Q^2$ we obtain:
$$F_{0\omega}(Q,\mu) = \Theta(Q-\mu) + \sum_{r=1}^{\infty}\int_{\mu^{2}}
^{Q^{2}}\prod_{i=1}^{r}\frac{d^{2}\mbox{\boldmath $q$}_{i}}{\pi q_{i}^{2}}
dz_{i}\frac{\bar{\alpha}_{S}}{z_{i}}z_{i}^{\omega}\Delta_{R}(z_{i},k_{i}) 
\nonumber\\
= 1 + \sum_{r=1}^{\infty}\sum_{n=r}^{\infty}C^{(r)}_{0}(n;T)\frac{\bar
\alpha_{S}^{n}}{\omega^{n}},$$
with $T=\ln(Q/\mu)$.

   Modifying the BFKL formalism to account for coherence we get the `CCFM'
~\cite{C} expression for 
$F_{0\omega}(Q,\mu)$:
$$F_{\omega}(Q,\mu) = \Theta(Q-\mu) + \sum_{r=1}^{\infty}\int_{0}^{Q^{2}}
\prod_{i=1}^{r}\frac{d^{2}\mbox{\boldmath $q$}_{i}}{\pi q_{i}^{2}}dz_{i}
\frac{\bar{\alpha}_{S}}{z_{i}}z_{i}^{\omega}\Delta(z_{i},q_{i},k_{i})
\Theta(q_{i}-z_{i-1}q_{i-1})$$
$$= 1 + \sum_{r=1}^{\infty}\sum_{n=r}^{\infty}\sum_{m=1}^{n}C^{(r)}(n,m;T)
\frac{\bar\alpha_{S}^{n}}{\omega^{2n-m}},$$
where we introduced the coherence improved 
Regge factor $\Delta(z_{i},q_{i},k_{i})=\exp\left[-\bar\alpha_{S}\ln\frac{1}
{z_{i}}\ln
\frac{k_{i}^{2}}{z_{i}q_{i}^{2}}\right];~ ~ k_{i} > q_{i},$ and for the 
first emission we take $q_{0}z_{0} = \mu$.

   In the formalism with coherence no collinear cutoff is needed, except on 
the emission of the first gluon. This is because subsequent collinear 
emissions are regulated by the angular ordering constraint and it is those 
collinear emissions which induce the additional powers of $1/\omega$. 
 In inclusive quantities the 
collinear singularities cancel. At a less inclusive level, such as for the 
associated distributions, the collinear singular terms need not cancel any 
more $~\cite{M}$. 

{\large \bf \boldmath 2 Equivalence of BFKL and CCFM approaches at order 
${\bar \alpha}_{S}^{3}$.}

The rates for emission 
of fixed numbers of resolved final-state gluons, together with any number 
of unresolvable ones, were calculated in Ref.$~\cite{jeffagustin}$
in the leading logarithmic approximation, to third order in $\bar\alpha_S$.
By resolved we mean having a transverse momentum larger 
than a resolution scale $\mu_{R}$ constrained by the collinear cutoff and 
the hard scale, $\mu \ll \mu_{R}\ll Q$. Within the leading log$(1/x)$ 
approximation,
 the resolved gluons can be 
identified as jets since any corrections arising from
additional radiation are suppressed by ${\cal O}(\alpha_s)$.

  To calculate a given $n$-jet rate we consider all the graphs with $n$ 
resolved gluons and any number of unresolvable ones.
We expand the Regge factors to order ${\bar \alpha}_{S}^{3}$. Doing so we 
find that the jet rates expressions calculated both in the multi-Regge (BFKL) 
approach and in the coherent (CCFM) approach are the same. Namely:

$$``{\rm 0-jet}" = \frac{(2\bar{\alpha}_{S})}{\omega}S
+\frac{(2\bar{\alpha}_{S})^{2}}{\omega^{2}}\left[\frac{S^{2}}{2}\right]
+\frac{(2\bar{\alpha}_{S})^{3}}{\omega^{3}}\left[\frac{S^{3}}{6}\right]
+...,$$   

$$``{\rm 1-jet}" = \frac{(2\bar{\alpha}_{S})}{\omega}T+\frac{(2\bar{\alpha}
_{S})^{2}}
{\omega^{2}}\left[TS-\frac{1}{2}T^{2}\right]+\frac{(2\bar{\alpha}_{S})^{3}}
{\omega^{3}}\left[\frac{1}{3}T^{3}-\frac{1}{2}T^{2}S+\frac{1}{2}TS^{2}\right]
+...,$$

$$``{\rm 2-jet}"= \frac{(2\bar{\alpha}_{S})^{2}}{\omega^{2}}
\left[T^{2}\right]+\frac{(2\bar{\alpha}_{S})^{3}}{\omega^{3}}
\left[T^{2}S-\frac{7}{6}T^{3}\right]+...,$$

$$``{\rm 3-jet}"=\frac{(2\bar{\alpha}_{S})^{3}}
{\omega^{3}}\left[T^{3}\right]+...,$$
with $T=\ln(Q/\mu_R)$ and $S=\ln(\mu_R/\mu)$. Note that in the calculation
 with coherence there exist stronger singularities ($\omega \rightarrow 0$) 
than occur in the BFKL approach but these additional ``coherence induced'' 
logarithms cancel when we sum all the graphs contributing to the jet rates 
and the final results are identical to those obtained without coherence. We 
will see now how this cancellation persists for $n$-jet rates to all 
orders in ${\bar \alpha}_{S}$.

{\large \bf \boldmath 3 Equivalence of BFKL and CCFM approaches at all orders
 in ${\bar \alpha}_{S}$.}

In a recent paper$~\cite{webber}$ the work of Ref.$~\cite{jeffagustin}$
was extended to all
orders, for any number of resolved gluons. The BFKL and CCFM formulations
were shown to give the same jet rates in leading logarithmic approximation 
to all
orders. The factorization of collinear singularities was demonstrated,
and a simple generating function for the jet multiplicity distribution
was obtained.

  Working within the multi-Regge (BFKL) approach, a simple 
expression for the $r$-jet rate was found:
$$R^{(r\res)}_\om(Q,\mR) = \frac{F^{(r\res)}_\om(Q,\mR,\mu)}{F_\om(Q,\mu)}
\>=\>\frac{1}{r!}
\left.\frac{\partial^r}{\partial u^r}\left(\exp\left(-\frac{2\alb}{\om}T\right)
\left[1+(1-u)\frac{2\alb}{\om}T\right]^{\frac{u}{1-u}}\;\right)\right|
_{u=0}\;.$$

   The same jet-rate generating function is
obtained taking account of coherence from the CCFM formulation of small-$x$ 
dynamics.
After convolution with the measured gluon structure function,
it gives the predicted jet rates in the leading logarithmic region
$\ln(1/x)\gg T=\ln(Q/\mR)\gg 1$, to all orders in $\as$, proving that the 
CCFM results for the multijet
contributions are equal to the BFKL predictions,
thus completing the all-orders 
extension of the results of Ref.~\cite{jeffagustin}. 

   From this generating function we can obtain quantities to all orders, for 
example the mean number
of jets and the mean square fluctuation in this number,
$$\langle r\rangle =
\left.\frac{\partial}{\partial u}R_\om(u,T)\right|_{u=1}\>=\>
\frac{2\alb}{\om}T +\frac{1}{2} \left(\frac{2\alb}{\om}T\right)^2\; ,~~~~~ 
\langle r^2\rangle -\langle r\rangle^2 = \frac{2\alb}{\om}T
+\frac{3}{2} \left(\frac{2\alb}{\om}T\right)^2
+\frac{2}{3}\left(\frac{2\alb}{\om}T\right)^3\;.$$
In general, the $p$th central moment of the jet multiplicity
distribution is a polynomial in $\alb T/\om$ of degree $2p-1$,
indicating that the distribution becomes relatively narrow in
the limit of very small $x$ and large $Q/\mR$.

{\large \bf \boldmath 4 Conclusions.}

 It has been shown that at leading logarithmic level the BFKL and CCFM 
approaches give
the same results for multi-jet rates in DIS at small $x$. Both methods 
give the same simple
 generating function for the jet rates. We will expect to find differences 
only at the sub-leading level and
in more differential quantities such as multi-jet rapidity correlations
~\cite{CMW}.

{\large \bf \boldmath 5 Acknowledgements.}

ASV is grateful to the organizing committee for the financial support of
 his participation in the Workshop.

\end{document}